# Burst Mode Ultrafast Laser Welding of Sapphire and Fe-36Ni Alloy with Non-optical Contact Condition

Yu Wang[1, 2], Nan Li[2, *], Yuxuan Li[2, 3], Yitong Chen[2], Qingwei Zhang[2], Jianing Zhao[2], Zhe Lin[2], Zihui Dong[2], Guochang Jiang[2], Zhengqiang Zhu[1, *], Shanglu Yang[2, 3, *]

[a] Jiangxi Key Laboratory of Intelligent Robot, College of Advanced Manufacturing, Nanchang University, Nanchang 330031, China.
[b] Shanghai Institute of Optics and Fine Mechanics, Chinese Academy of Sciences, Shanghai 201800, China.
[c] Center of Materials Science and Optoelectronics Engineering, University of Chinese Academy of Sciences, Beijing 100049, China.

**Abstract**
Ultrafast laser welding provides a promising approach for high precision integration of transparent and metallic materials. However, its practical application remains constrained by the precise regulation of the interfacial gap. This study investigates the interfacial response and bonding mechanism of sapphire and Fe-36Ni alloy joints under controlled non-optical contact conditions using burst mode ultrafast laser irradiation. A polymer interlayer was introduced between naturally stacked samples to establish a variable interfacial gap, allowing systematic evaluation of gap-dependent morphology, melting behavior, and elemental transport. By redistributing the pulse energy into sequential sub-pulses, the burst mode reconstructs the temporal energy-deposition process, yielding enhanced plasma–material coupling and stable thermal accumulation. Compared with single pulse irradiation, burst mode sustains continuous bonding across gaps exceeding 10 μm—far beyond the failure threshold of the single pulse mode—and forms a fusion zone 82% larger. Fracture surface and cross-sectional analyses of SEM and EDS results confirm that sequential sub-pulses promote extensive sapphire melting, droplet-driven gap bridging, and enhanced Al–Fe interdiffusion at the interface. These results provide a scientific basis for high-gap-tolerance ultrafast laser welding and scalable integration of transparent–metal hybrid components in advanced optoelectronic and precision engineering applications.

## 1. Introduction

Highly integrated connecting components of transparent optoelectronic materials and metals are widely utilized in fields such as microelectronic packaging, optoelectronic devices, and advanced manufacturing [[1]–[4]]. In these aspects, owing to extremely short pulse duration, high peak power, and nonlinear absorption mechanisms [[5]–[9]], ultrafast lasers have gradually replaced conventional methods and emerged as a powerful tool for achieving glass–metal bonding without intermediate layers, offering minimal thermal influence and high spatial precision [[10],[11]]. In light of the increasing emphasis on green manufacturing and high-precision assembly, this technology is now recognized as an efficient, clean, and scalable solution [[12]].

Previous studies have demonstrated that the interfacial gap is one of the key factors limiting the welding performance of ultrafast lasers. Cvecek et al. [[13]] reported that intimate contact is essential to achieve high strengths; when the gap exceeds 130 nm, defects inside the welded seam tend to generated., Richter et al. [[14]] introduced the interfacial gap is usually strictly controlled within the submicron range (100–500nm) to prevent the escape of plasma during the interaction process. Our previous study [[15]] further showed that micron-scale gap variation triggers distinct melting–spallation–interlocking regimes, emphasizing the decisive role of gap-dependent interfacial physics. Under the optical contact condition (gap ⩽150nm) achieved through precise surface preparation and application of appropriate preload, stable localized melting and high-quality joint between transparent materials and metals can be realized, and the interfacial interaction mechanisms have been progressively clarified [[16]]–[[19]]. However, in practical engineering applications, component surfaces often exhibit high roughness or complex curvature, making it extremely challenging and time-consuming to achieve the required surface quality via mechanical polishing. Moreover, the mechanical pressure applied to minimize the gap will introduce residual stress and potential microcracks defects inside the material, adversely affecting joint integrity [[20]]. Consequently, there is an urgent need to explore a welding process that accommodates realistic interfacial conditions to enable reliable glass-metal bonding.

In recent years, burst mode has shown great potential in overcoming this bottleneck [[21]]–[[23]]. By introducing multiple sub-pulses with high repetition rates (MHz or GHZ) within a single pulse envelope, this technique reshapes the temporal structure of energy deposition. This allows subsequent sub-pulses to temporally overlap with the plasma plume and thermal field induced by preceding sub-pulses, resulting in more efficient and controllable energy coupling [[24],[25]]. Burst mode has demonstrated remarkable effectiveness in material removal [[26],[27]], surface modification [[28]], and micro/nanostructure fabrication [[29],[30]]. For glass-metal welding systems, compared with the single pulse mode, burst mode not only increase plasma density and absorption depth but also enable temporal regulation of electron-lattice energy transfer and local thermal field evolution. It offers a potential for reliable welding under non-ideal gap conditions.

In this study, a variable-gap model was constructed by introducing a 10μm polymer spacer layer between naturally stacked sapphire and Fe-36Ni alloy substrates. This configuration enables a controlled investigation of interfacial bonding behavior under non-optical contact condition. By directly comparing the joint morphologies obtained under single pulse mode and burst mode across varying gap widths, we demonstrate the superior joining capability of the burst regime. Through precise modulation of the sub-pulse sequences and energy deposition rate, the interfacial thermal field and localized melting behavior are effectively controlled, enabling direct welding of sapphire and Fe-36Ni alloy across interfacial gap exceeding 10μm. This work elucidates the underlying physical mechanisms of gap bridging in burst mode ultrafast laser welding and provides a quantitative foundation for the development of high-tolerance, non-contact heterogeneous joining strategies.

## 2. Experimental setup and methods

A picosecond laser (Amber NX IR-50, Bellin Laser) combine with galvo scanner and F-theta lens (f=100mm) was employed for the experiments (Fig. 1(a)), the focal spot diameter is 40 μm. Conduct process experiments using both single pulse mode ($N$=1) and burst mode ($N$>1) of the laser (Fig. 1(e)), the number of sub-pulse $N$ varied from 1 to 15. All of laser parameters are shown in Table. 1.

The sapphire used in this study were double-side polished (Sa = 0.28 nm, Sz = 9.27 nm) with dimensions of 10×10×3 mm$^3$, and the Fe-36Ni alloy substrates were single-side polished (Sa = 27.52 nm and Sz = 610.9 nm) with dimensions of 10×10×2 mm$^3$. In order to simulate welding behavior under different interfacial gap conditions, a 10 μm polymer film was inserted between the two materials, allowing a continuously varying gap distribution along the interface. Subsequently, the interface was secured on the opposite side by a welded line, achieving stable gap distribution, as shown in Fig. 1(c). The interfacial gap was measured using a customized white light interferometer (METAFILN-WO25, LightE-Technology Co. Ltd.) shown in Fig. 1(b, f), and the statistical results are presented in Fig. 1(d). The variation of the gap with respect to the measurement position can be fitted with a linear function:

$$y = 1.85x + 3.2 \quad (1)$$

Where $x$ is the distance from the welded line (mm), $y$ is the gap thickness $d$ (μm). By subsequently processing single line paths along the direction of the varying gap, welded seams with a continuous gradient in morphology were obtained, enabling systematic evaluation of interfacial behavior under different gap conditions.

After welding, the macro morphology of the joints was observed using a 3D digital microscope (VHX-600, Keyence). Microstructural and compositional analyses of the gold-coated cross-sectional and fracture surfaces were conducted via field-emission scanning electron microscopy (SEM5000, CIQTEK Co. Ltd.) equipped with energy-dispersive X-ray spectroscopy (Ultim Max40, Oxford Instruments).

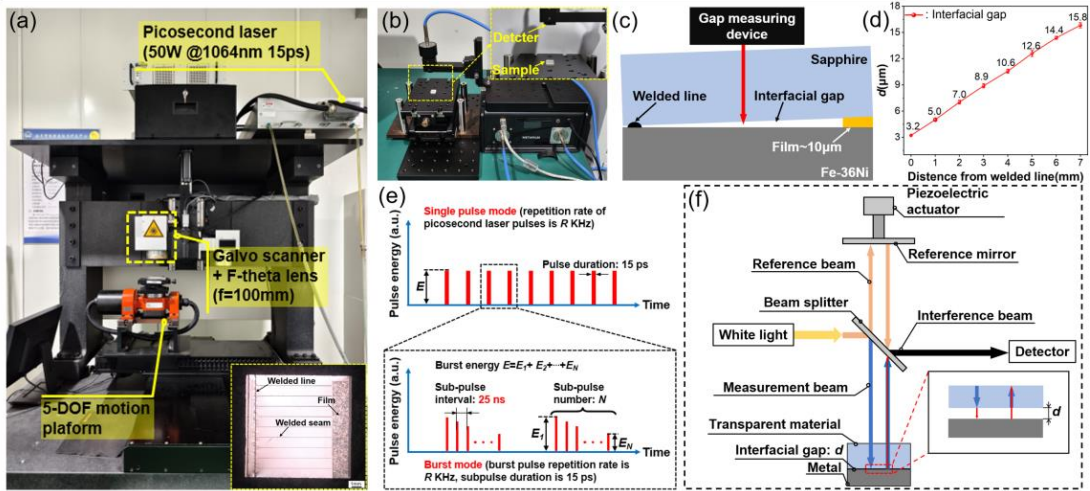

Fig. 1 (a) Picosecond laser welding system and welded specimen of sapphire/Fe-36Ni alloy; (b) interfacial gap measurement setup; (c) schematic diagram of the interfacial gap regulation strategy; (d) variation in the interfacial gap; (e) schematic diagram of burst mode principle; (f) schematic diagram of the interfacial gap measurement principle using white-light interferometry.

Table. 1 Laser parameters of ultrafast laser welding sapphire and Fe-36Ni alloy

| Laser parameter | Value |
| --- | --- |
| Central wavelength (nm) | 1064 |
| Pulse width (ps) | 15 |
| Repetition rate (KHz) | 1000 |
| Pulse energy (μJ) | 22 |
| Scanning speed (mm/s) | 100 |
| Sub-pulse number: $N$ | 1/3/5/7/9/15 |
| Sub-pulse interval time (ns) | 25 |

## 3. Results and discussion

### 3.1. Influence of sub-pulse number on fracture morphology

Under small initial gaps, the metal-side fracture morphology evolves systematically with increasing sub-pulse numbers (Fig. 2($a_1$-$f_1$)). At low $N$ (1, 3), microcracks within the sapphire propagate rapidly transversely along the crystal orientation under stress, giving rise to flake-like spallation and typical cleavage steps (Fig. 2($a_1$, $b_1$)). This arises from the very high peak power of a single pulse or few sub-pulses, and the geometrical confinement of the narrow gap, which promotes laser-matter coupling. The laser energy is instantaneously deposited through nonlinear absorption and electronic excitation, dissipating before electron-phonon thermal relaxation and thus suppressing notable sapphire softening or melting.

As $N$ increases (5, 7, 9), the energy of each sub-pulse decreases, resulting in a gentler excitation and heating process. Subsequent sub-pulses deposit energy into the existing plasma via inverse bremsstrahlung, leading to a gradual rise in local temperature and a pronounced thermal accumulation. This cumulative energy input facilitates localized melting and solidification of the joint, thereby weakening the brittle cleavage characteristics of the fracture. Meanwhile, the repeated disturbance of the molten pool by successive sub-pulses within the confined space ejects a small amount of molten material, forming burr-like spatter structures (Fig. 2($d_1$, $e_1$)). The statistical variation of weld width ($W_1$) with respect to $N$ is shown in Fig. 6(a), $W_1$ initially increases and then decreases as $N$ increases, reaching a maximum at $N=3$. Notably, the standard deviation is largest at $N=1$ reflects the energy coupling and mass transport processes under single pulse irradiation exhibit strong randomness. In contrast, the burst mode achieves more stable thermal field and molten pool evolution over time, improving the consistency of weld formation and process controllability.

The performances of the two modes under large gap conditions ($d$ ~16 μm) are shown in Fig. 2($a_2$, $f_2$). The burst mode achieves stable gap-bridging joining, while the single pulse mode fails to establish an effective bond. In the single pulse mode ($N=1$), the high-peak pulse is absorbed by the metal surface, where instantaneous evaporation and intense recoil pressure drive the ejection of metal vapor. The large gap allows free plume expansion, causing the ejection of metallic particles that subsequently deposit along the molten pool boundary. Central ablation holes formed due to thermal shock, the molten metal is squeezed to both sides and accumulates under the combined effects of high-pressure vapor and backflow, eventually solidifying into a dam-like resolidified structure. With 3-7 sub-pulses, burst mode distribute energy over multiple deposition events, mitigating the ablation effect and promoting continuous sapphire melting, part of the molten material bridges the gap, giving rise to droplet-like bridging morphologies.

The distribution width of micro/nanoparticles($W_2$) under large gap conditions was statistically analyzed (Fig. 6(a)). When $N=1$, the strongest laser impact on the metal molten pool produces a maximum $W_2$ of ~200 μm. For $N>1$, $W_2$ drops sharply to ~100 μm and decreases gradually thereafter. However, when $N$ reaches 15, no melting occurs at the interface, $W_2$ reduces to zero. The effective energy reaching the interface is insufficient to sustain cross-interface melting, and only laser-induced periodic surface structures (LIPSS) can be induced on the metal surface—its energy level is much lower than the melting threshold of sapphire. The laser-material interaction mechanism shifts to a near-threshold metal surface modification process.

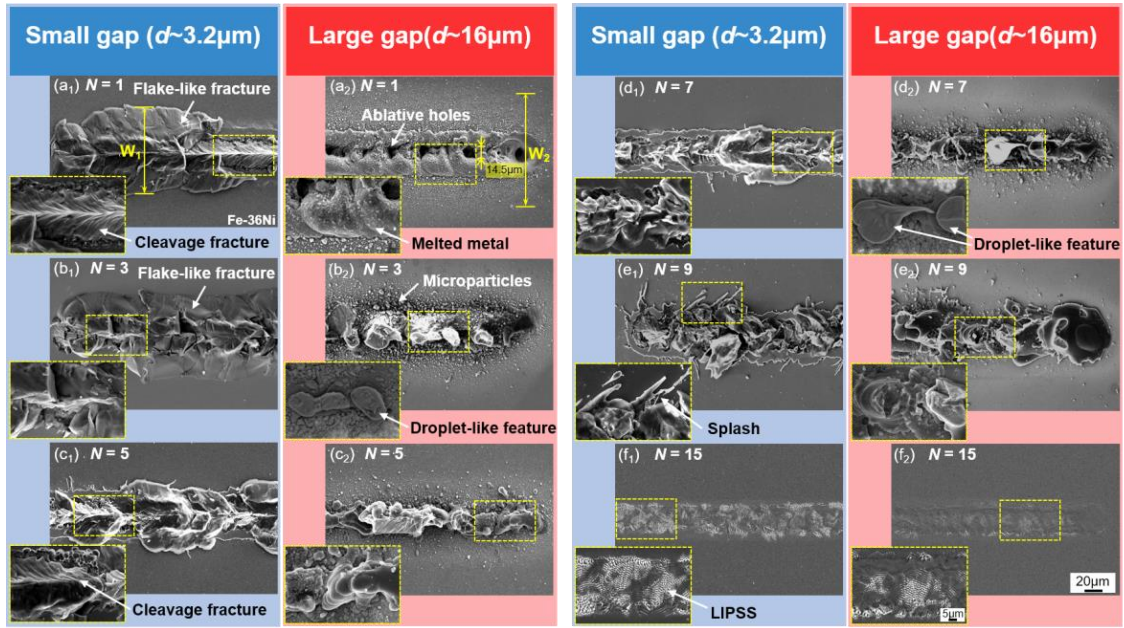

Fig. 2 Fracture surface morphologies at both ends of welded seam on Fe-36Ni side processed by ($a_1$, $a_2$) single pulse mode ($N$=1) and ($b_1$-$f_1$, $b_2$-$f_2$) burst mode with different sub-pulses ($N$=3, 5, 7, 9, 15).

### 3.2. Weld morphology evolution with increasing interfacial gap

The complete weld morphology and the evolution of its local features with gap variation (Fig. 1(d)) under two modes are shown in Fig. 3. In the single pulse mode, as the gap increases, the sapphire rapidly transitions from continuous cleavage fracture to a discontinuous fragmented fracture surface, as indicated by the yellow dashed line in Fig. 3 (b) ($d$ ~3.6 μm). Residual bulk sapphire finally be observed at gap of ~6 μm (Fig. 3 (c)). With interfacial gap increases further, molten metal particles are distributed on both sides of the welded seam, and their area gradually expands. The welded seam is then composed of continuous ablation holes and protruding resolidified metal, with the sapphire completely disappearing (Fig. 3(d)).

In the burst mode, the weld morphology exhibits a two-stage evolution with increasing interfacial gap. At a gap of ~ 6.9 μm (marked by the yellow dashed line in Fig. 3 (g)), a relatively continuous fracture structure is maintained. Beyond this point, the interface transforms into a discontinuous droplet-like bridging morphology accompanied by sapphire particles.

This droplet-bridging feature persists up to the maximum experimental gap of 16 μm, which indicates that multi-level sub-pulses can stably sustain interface melting and migrating of sapphire over a wider gap range. The preceding sub-pulses induce local plasma and an initial molten layer, while the subsequent sub-pulses significantly enhance the effective energy density at the interface through the combined effects of thermal accumulation and suppression of energy scattering. Meanwhile, the pulse peak power is significantly reduced, preventing energy loss caused by severe ablation and intense plume impact.

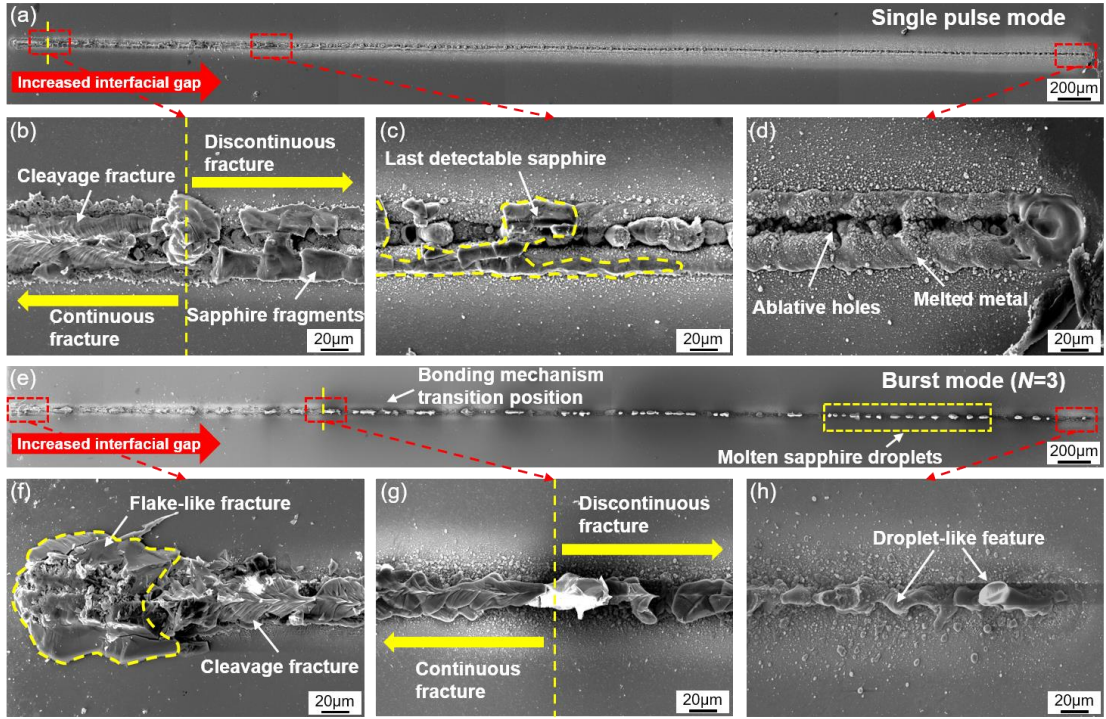

Fig. 3 Overall fracture surface characteristics of welded seam on Fe-36Ni alloy side with varying interfacial gaps processed by (a) single pulse mode and (b) burst mode ($N$=3); (b-d) are enlarged images in (a) and (f-h) are enlarged images in (e).

3.3 Interfacial microstructure and bonding mechanism

Fig. 4 and Fig. 5 show the typical fracture morphology at the maximum stably achievable gap (~16 μm), and the weld cross-sectional morphology at a 10 μm gap, along with the surface scanning results of major elements. Together, these results reveal significant spatial differences in interfacial energy deposition and material response between the two temporal regulation modes.

Under single pulse irradiation, the Al element in Fig. 4(a) is discretely distributed on the metal substrate, with slight enrichment only at the welded seam center. This indicates that these Al-enriched phases formed through local melting and re-solidification, detached from the lower surface of the sapphire, crossed the gap, and migrated to the metal side. These Al-enriched aggregates did not dissolve uniformly in the metal molten pool; instead, they solidified as granular aggregates. Combined with the cross-sectional images shown in Fig. 5(a), the migration process and molten pool dynamic can be further elucidated. Transient high-peak energy induces ultrafast heating and non-equilibrium electron-lattice coupling in local regions, leading to fragmentation and detachment of micro-regions on the sapphire surface. Driven by laser recoil pressure, these fragments enter the metal molten pool, which is manifested as a distinct truncation feature between the Al-enriched region on the metal side and the sapphire matrix.

Intense local heating establishes a steep temperature gradient in the molten pool, triggering Marangoni convection driven by the surface tension gradient [[31]]. This flow pattern, co-driven by Marangoni convection and recoil pressure, pushes the molten metal laterally away from the center of the focused laser beam. After approaching the molten pool boundary, the metal deflects and is further transported outward along the liquid-solid interface. Due to the lack of constraint from the overlying matrix under large gap conditions, the deflected metal fluid gradually extends to the free surface and accumulates locally under the combined action of continuous convective driving and material deposition (Fig. 6(d)). Finally, a ridge-like raised structure with a low center and high sides is formed during solidification.

Throughout the energy interaction process, sapphire and Fe-36Ni alloy exhibit independent energy absorption and material responses. This prevents the establishment of a bridging mechanism and fails to achieve effective closure of the interfacial gap. In contrast, joint processed with 3 sub-pulses display distinctly different micro-morphology evolution and elemental dynamics. The fracture morphology on the metal side and the surface scanning results of major elements (Al and Fe) is shown in Fig. 4(b).

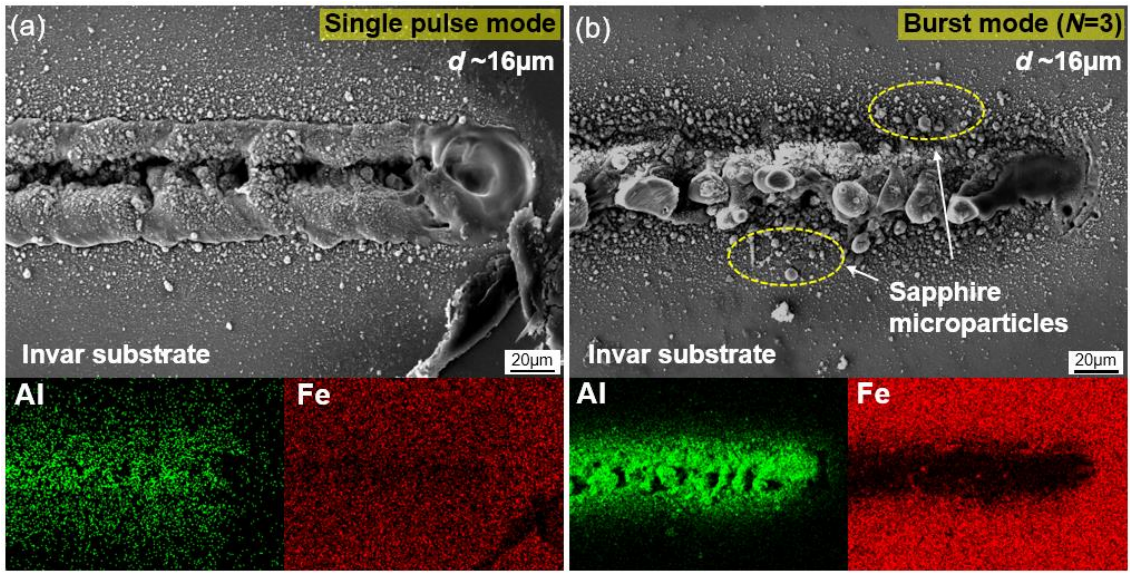

Fig. 4 Fracture surface characteristics of welded seam at large interfacial gap (~16μm) and the corresponding EDS elemental mappings of Al and Fe processed by (a) single pulse mode, (b) burst mode ($N$=3).

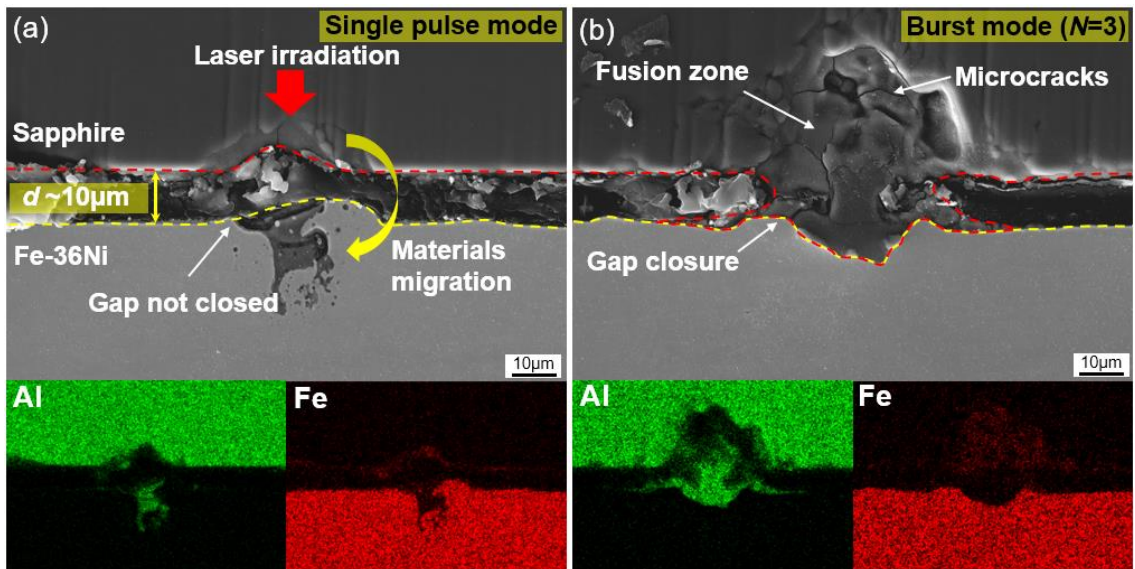

Fig. 5 SEM cross-sectional images of welded joints at 10 μm interfacial gap and the corresponding EDS elemental mappings of Al and Fe processed by (a) single pulse mode, (b) burst mode ($N$=3).

The three sub-pulses are incident sequentially at nanosecond intervals. Through stepwise energy deposition, they induce linear absorption in sapphire, resulting in complete melting and extension toward the metal surface in the form of droplets. Thus, forming a bridging structure that spans the large gap interface. Simultaneously, the stirring effect of subsequent sub-pulses on the molten pool triggers localized droplet spatter, causing a small amount of molten sapphire to be distributed around the welded seam in the form of microparticles. The Al element is highly enriched in the weld region, with a slight protrusion at the welding center. This indicates that during fracture process, the sapphire region within the joining gap—acting as the weak link in joint strength—failed first.

Under this gap-bridging mechanism dominated by the dynamic behavior of molten glass [[32],[33]], a large-area fusion zone is formed on the sapphire side. The molten sapphire jet ejects downward to fill the interfacial gap, achieving sufficient contact and wetting with the metal surface layer, effectively closing the originally existing gap, as shown in Fig. 5(b). EDS elemental mapping results reveal significant diffusion within the fusion zone: Fe diffuses upward toward the top of the fusion zone, while Al migrates downward and spreads laterally along the region close to the metal side. This reciprocal diffusion not only alleviates the abrupt compositional transition across the interface but also provides favorable chemical contact conditions, promoting the realization of metallurgical bonding. Microcracks are observed on the cross-section, attributed to the intense thermal stresses induced by laser energy input. The positions of these microcracks are highly consistent with the cleavage regions identified in the fracture analysis, serving as the main factor limiting joint strength of the burst mode (Fig. 6(e)).

The large-area joint obtained under the burst mode is shown in Fig. 6(c). A thin film with a central circular aperture was inserted between the sapphire and Fe-36Ni alloy as a gap-control layer, enabling uniform regulation of the interfacial spacing across an area slightly smaller than the sample diameter and maintaining a stable gap of approximately 10 μm. As a result, a structurally continuous and morphologically uniform bonding interface was achieved across the entire 4 × 4 mm² processed region. These results demonstrate that, under optimized burst modulation, ultrafast laser irradiation can realize dense and stable sapphire–metal joining even under ten-micrometer-scale interfacial gap.

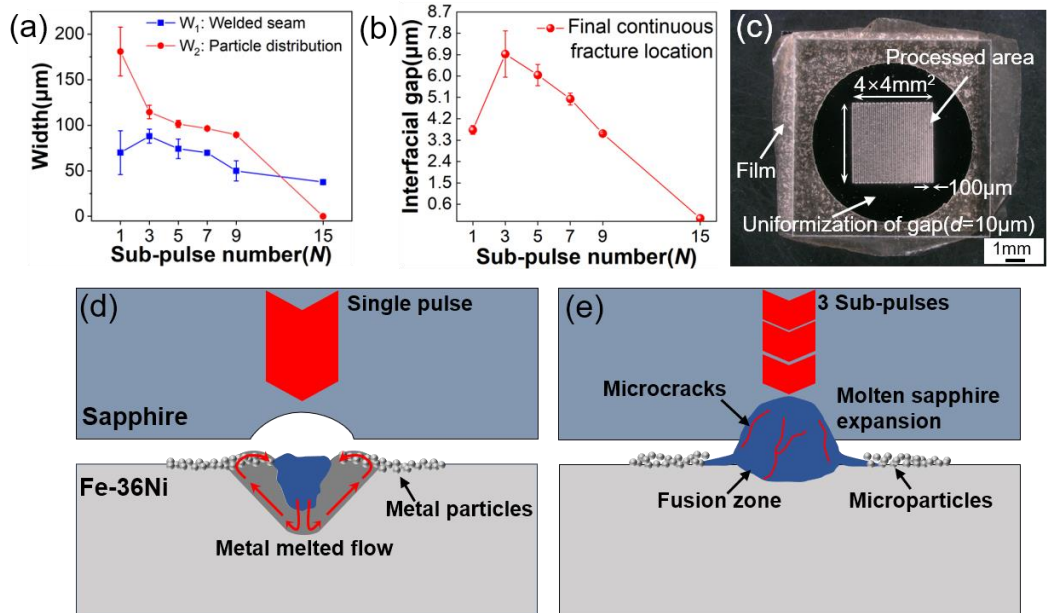

Fig. 6 (a) Evolution of the welded seams under optical contact ($W_1$) and the particles distribution under non-optical contact ($W_2$) as a function of sub-pulse number in Fig 2; (b) Variation of interfacial gap thickness at the final continuous fracture location for welded seams as a function of sub-pulse number; (c) large-area joined specimen with a uniformly controlled 10 μm interfacial gap; schematic diagram of the bonding mechanism: (d) single pulse mode, (e) burst mode.

## 4. Conclusions

This study comparatively analyzed the joint formation mechanism and interfacial evolution of ultrafast laser single pulse mode and burst mode in welding of sapphire and Fe-36Ni alloy. It revealed the advantages and underlying mechanism of burst mode welding under non-ideal assembly conditions. The main conclusions are as follows:

**(1) Interfacial energy coupling behavior:** By temporally distributing laser energy into multiple sub-pulses, the burst mode effectively suppresses instantaneous ablation and enhances the stability of plasma–material interaction, achieving higher thermal accumulation and controlled melting across the sapphire–metal interface. Optimal performance is achieved at a sub-pulse number of three.

**(2) Gap dependence of weld morphology:** In the burst mode, as interfacial gap increases to ~6.9 μm, the joint transitions from a continuous cleavage to a droplet-like bridging morphology. Stable sapphire melting and migration toward the metal side are maintained across gap of 16 μm. In the single pulse mode, stable melting cannot be sustained once the gap surpasses ~3.6 μm, leading to fragmented fracture surfaces; when the gap exceeds ~6 μm, sapphire melting ceases completely, leaving only ablated holes and resolidified metal.

**(3) Interfacial gap-bridging mechanism:** The heterogeneous interface bonding process at a 10 μm gap under the burst mode is dominated by the dynamic behavior of molten sapphire. Sequential sub-pulses induce progressive sapphire melting and droplet transfer across the interface, forming a fusion zone 82% larger than that produced by single pulse irradiation. This mechanism allows efficient material transport and interfacial closure across wide gaps.

**(4) Broader elemental interdiffusion:** During burst mode irradiation, interfacial melting promotes bidirectional diffusion, which facilitates the formation of a heterogeneous mixing layer that mitigates abrupt compositional transitions and strengthens metallurgical bonding between sapphire and Fe-36Ni alloy.

**FUNDING**
National Key Research and Development Program of China (2023YFB4605503)**AUTHOR DECLARATIONS**
Conflict of Interest
The authors have no conflicts to disclose.**DATA AVAILABILITY**
The data that support the findings of this study are available from the corresponding author upon reasonable request.